\documentclass[12pt,preprint,prb]{revtex4}
\usepackage{graphicx}
\usepackage{psfrag}
\usepackage{amssymb,amsfonts,amsmath}

\newcommand{\lbe}[1]{\begin{equation} \label{#1}}
\newcommand{\ee}{\end{equation}}
\newcommand{\dd}{\mathrm{d}}
\newcommand{\e}{\mathrm{e}}

\begin{document}

\title{An analytic cylindrically symmetric solution for collapsing dust}
\author{J. Hennig}
\email[e-mail: ]{J.Hennig@tpi.uni-jena.de}
\author{G. Neugebauer} 
\affiliation{Theoretisch-Physikalisches Institut, Friedrich-Schiller-Universit\"at Jena, 
Max-Wien-Platz 1, 07743 Jena, 
Germany}

\begin{abstract}
Dust configurations are the simplest models for astrophysical
objects. Here we examine the gravitational collapse of an infinite
cylinder of dust and give an analytic interior solution. 
Surprisingly, starting with a cylindrically symmetric ansatz one arrives
at a 3-space with constant curvature,
i.e. the resulting metric describes a piece of the Friedman interior of
the Oppenheimer-Snyder collapse. Indeed, by introducing double polar
coordinates, a 3-space of constant curvature can be interpreted as a
cylindrically symmetric space as well. This result shows afresh that
topology is not fixed by the Einstein equations.

\end{abstract}
\maketitle

%%%%%%%%%%%%%%%%%%%%%%%%%%%%%%%%%%%%%%%%%%%%%%%%%%%%%%%%%%%%%%%%%%%%%%%%%%%%%
\section{Introduction}
Astrophysical collapse processes are one of the main sources of
gravitational radiation. However, the complexity of the
Einstein equations makes it very difficult to model such processes
mathematically. This is why we turn here to a dust model. Dust is
the simplest material for astrophysical and cosmological studies and may
be interpreted as a many-particle system with only gravitational
interaction between its particles.  

Unfortunately, one cannot study gravitational radiation in spherically
symmetric spacetimes (e.g. the Oppenheimer-Snyder collapse
\cite{Oppenheimer}) since the Birkhoff theorem shows radiation to be absent. For this
reason, much work has been devoted to studying cylindrical symmetry, e.g.
\cite{Z1,Z2,Z3,Z4,Z5}.

There is an interesting suggestion for the description of cylindrically symmetric
dust models by Singh and Vaz. In \cite{Singh} they examine the collapse of
an infinite cylinder of dust making the assumption that the axial and
azimuthal metric functions are connected by a simple relation.
The authors discuss the resulting field equations qualitatively and note
a striking analog to the Friedman equations that describe the interior
of the Oppenheimer-Snyder collapse.

This paper is meant to show that this statement is not accidental. By
\emph{solving} the differential equations we obtain a geometry that is a
piece of the Friedman Universe (with a 3-space of constant
curvature). Nevertheless, by introducing double polar coordinates the
geometry takes the form of a cylindrically symmetric space-time as well.

%%%%%%%%%%%%%%%%%%%%%%%%%%%%%%%%%%%%%%%%%%%%%%%%%%%%%%%%%%%%%%%%%%%%%%%%%%%%%

\section{Cylindrically symmetric line element}
We want to find solutions of the Einstein equations which describe an
infinite cylinder of dust (no pressure: $p=0$). In this case, the
stress-energy tensor of an ideal fluid simplifies to
\begin{equation}\label{2.1}
T_{ij}=\mu(\rho,t)u_i u_j\ ,
\end{equation}
where $\rho$ and $t$ are the radial and the time component of the
co-moving coordinates $(\rho,\varphi,z,t)$ and $u_i$ is the
four-velocity with the special structure
\begin{equation}\label{2.2}
(u^i)=(0,0,0,1)
\end{equation}
(we choose units in which $G=c=1$).
We discuss the general cylindrically symmetric line element
\begin{equation}\label{2.3}
\dd s^2=L^2(\rho,t)\,\dd
\rho^2+B^2(\rho,t)\,\dd\varphi^2+M^2(\rho,t)\,\dd z^2-\dd t^2\ ,
\end{equation}
but with the additional assumption that for the functions $B$ and
$M$ of the line element the condition
\begin{equation}\label{2.3a}
B=\rho M
\end{equation}
holds. This
assumption generalizes a property of the surface geometry of cylinders
in flat 3-spaces: Surfaces $\rho=\textrm{constant}$, $t=\textrm{constant}$ have the  line
element $\dd s^2=M^2(\rho^2\dd\varphi^2+\dd z^2)$ which differs only by
a conformal factor (a constant for every slice) from the usual
expression $\dd s^2=\rho^2\dd\varphi^2+\dd z^2$.

%%%%%%%%%%%%%%%%%%%%%%%%%%%%%%%%%%%%%%%%%%%%%%%%%%%%%%%%%%%%%%%%%%%%%%%%%%%%%

\section{Solution of the field equations}

From ${T^{ij}}_{;j}=0$ and by using eq. \eqref{2.1} one finds
\begin{equation}\label{2.4}
(\mu u^i)_{;i}=\frac{1}{\sqrt{-g}}(\mu
u^i\sqrt{-g})_{,i}=\frac{1}{\sqrt{-g}}(\rho L M^2\mu)_{,t}=0
\end{equation}
with the solution
\begin{equation}\label{2.5}
\mu(\rho,t)=\frac{2\psi(\rho)}{\kappa\rho LM^2}\ ,
\end{equation}
where $\kappa$ is Einstein's gravitational constant, $\kappa=8\pi$.
Using this result, the Einstein equations are (see \cite{Singh})
\begin{equation}\label{2.6}
\frac{2\ddot B}{B}+\frac{\dot
B^2}{B^2}=\frac{1}{L^2}\left(\frac{B'^2}{B^2}-\frac{B'}{\rho B}\right)\ ,
\end{equation}
\begin{equation}\label{2.7}
\frac{2\ddot B}{B}+\frac{\ddot L}{L}=-\frac{\psi(\rho)}{\rho LM^2}\ ,
\end{equation}
\begin{equation}\label{2.8}
\frac{2\dot M'}{M}+\frac{\dot M}{\rho M}=\frac{\dot
L}{L}\left(\frac{2M'}{M}+\frac{1}{\rho}\right)\ ,
\end{equation}
\begin{equation}\label{2.9}
\frac{2M'}{M}=\frac{L'}{L}\ ,
\end{equation}
where a dot and prime denote derivatives with respect to $t$ and $\rho$ respectively. 
Eq. \eqref{2.9} has the solution
\begin{equation}\label{2.10}
L(\rho,t)=h(t)M^2(\rho,t)
\end{equation}
and eq. \eqref{2.8} has the first integral
\begin{equation}\label{2.11}
(\sqrt{\rho}M)'=g(\rho)L(\rho,t)\ .
\end{equation}
The combination of these two relations gives
\begin{equation}\label{2.12}
(\sqrt{\rho}M)'=h(t)\frac{g(\rho)}{\rho}(\sqrt{\rho}M)^2
\end{equation}
This equation can be integrated. One obtains
\begin{equation}\label{2.13}
\frac{1}{\sqrt{\rho}M}=h(t)\Big[G(\rho)+d(t)\Big]\ ,
\end{equation}
where
\begin{equation}\label{2.13a}
G(\rho):=-\int\frac{g(\rho)}{\rho}\,\dd\rho\ .
\end{equation}
Using these results, the metrical coefficients take the form
\begin{equation}\label{2.14}
M=\frac{1}{h(t)\sqrt{\rho}[G(\rho)+d(t)]},\
B=\frac{\sqrt{\rho}}{h(t)[G(\rho)+d(t)]},\
L=\frac{1}{h(t)\rho[G(\rho)+d(t)]^2}\ .
\end{equation}

We have not yet ensured that equations \eqref{2.6} and \eqref{2.7} are fulfilled. 
If one substitutes $B$ and $L$ from \eqref{2.14} into eq. \eqref{2.7} then one
finds 
\begin{equation}\label{2.15}
-\rho\psi(\rho)=\left[\frac{3\dot H^2-3\ddot
H}{(G(\rho)+d(t))^4}+\frac{8\dot H\dot d-4\ddot
d}{(G(\rho)+d(t))^5}+\frac{10\dot d^2}{(G(\rho)+d(t))^6}
\right]\e^{-3H}
\end{equation}
with the definition $H(t):=\ln h(t)$.
The derivative with respect to time $t$ of this equation leads to 
$\dot d=0$. Hence one can set, without loss of generality,
\begin{equation}\label{2.17}
d\equiv 0\ ,
\end{equation}
because $d$ only appears in the form $G(\rho)+d$ and would thus merely
lead to a redefinition of the metric function $G$.
Then \eqref{2.15} simplifies to
\begin{equation}\label{2.18}
-\rho\psi(\rho)G^4(\rho)=(3\dot H^2-3\ddot
H)\e^{-3H}=-K_0=\textrm{constant}\ .
\end{equation}
That this expression is constant, follows from the fact that the
left-most expression is a function of $\rho$ alone, whereas the
expression to its right is a function $t$ alone. The mass density then
is given by
\begin{equation}\label{2.19}
\mu=\frac{2K_0}{\kappa}\e^{3H(t)}\ .
\end{equation}
Thus $\mu$ is a spatial constant for every time $t$. $H$ has to
satisfy the equation
\begin{equation}\label{2.20}
\ddot H-\dot H^2=\frac{K_0}{3}\e^{3H}\ .
\end{equation}
We have yet to satisfy eq. \eqref{2.6}. By using \eqref{2.14} with $d=0$
one finds
\begin{equation}\label{2.21}
-(2\ddot H-3\dot H^2)\e^{-2H}=\rho^2 G^2
{G'}^2-\frac{1}{4}G^4=K_1=\textrm{constant}\ .
\end{equation}
Again the left hand side is a function of another variable than the
right hand side and so both sides have to be constant. This leads to
another differential equation for $H$,
\begin{equation}\label{2.22}
\ddot H-\frac{3}{2}\dot H^2=-\frac{K_1}{2}\e^{2H}\ .
\end{equation}
Instead of the system of equations \eqref{2.20}, \eqref{2.22} one could
use the two combinations
\begin{equation}\label{2.23}
\ddot H=K_0\e^{3H}+K_1\e^{2H}\ ,\quad \dot
H^2=\frac{2}{3}K_0\e^{3H}+K_1\e^{2H}\ .
\end{equation}
The first equation in \eqref{2.23} follows from the second one by
differentiating with respect to time $t$ (for the case $\dot H \neq 0$, but the
case $H=\textrm{constant}$ is not possible, because then  one would
find $K_0=0$ from \eqref{2.20} and from this $\mu=0$ from \eqref{2.19}
--- i.e. a vacuum).

Due to the freedom of coordinate transformations, we need only study the cases $K_1=-\varepsilon$ with
$\varepsilon=0,\pm 1$. With the definitions $a(t):=\e^{-H(t)}=1/h(t)$
and $a_\textrm{m}=\frac{2}{3}K_0$, the second equation in \eqref{2.23} takes
the form
\begin{equation}\label{2.24}
\dot a^2-\frac{a_\textrm{m}}{a}+\varepsilon=0\ ,\quad\varepsilon=0,\pm
1\ .
\end{equation}
Surprisingly, the scale factor $a$ satisfies the dynamical equation of
the Friedman cosmology. Hence, its solutions are
\begin{align}
&\varepsilon=+1: & a(\eta)&=\frac{a_\textrm{m}}{2}(1-\cos\eta)\ ,\quad
t(\eta)=\frac{a_\textrm{m}}{2}(\eta-\sin\eta)\nonumber\\
&\varepsilon=0: &
a(t)&=\left(\frac{3}{2}\sqrt{a_\textrm{m}}t\right)^{2/3}\label{2.25}\\
&\varepsilon=-1: & a(\eta)&=\frac{a_\textrm{m}}{2}(\cosh\eta-1)\ ,\quad
t(\eta)=\frac{a_\textrm{m}}{2}(\sinh\eta-\eta)\ .\nonumber
\end{align}

Solving eq. \eqref{2.21} one obtains
\begin{align}
&\varepsilon=+1: & G^2(\rho)&=\frac{\rho}{\rho_0}+\frac{\rho_0}{\rho}\nonumber\\
&\varepsilon=0: &
G^2(\rho)&=\frac{\rho_0}{\rho}\label{2.26}\\
&\varepsilon=-1: & G^2(\rho)&=\left|\frac{\rho}{\rho_0}-\frac{\rho_0}{\rho}\right|\nonumber
\end{align}
with an integration constant $\rho_0>0$. Finally, with the substitution
$\rho=\rho_0\tan \chi$ (in the case $\varepsilon=+1$) or $\rho=\rho_0\tanh\chi$ (in
the case
$\varepsilon=-1$) one arrives (after stretching the coordinates to
eliminate the constant $\rho_0$) at the line elements
\begin{align}
&\varepsilon=+1: & \dd
s^2&=a^2(t)\left(\dd\chi^2+\sin^2\!\chi\,\dd\varphi^2+\cos^2\!\chi\,\dd
z^2\right)-\dd t^2\nonumber\\[1.5ex]
&\varepsilon=0: & \dd s^2 & =
a^2(t)\left(\dd\rho^2+\rho^2\dd\varphi^2+\dd z^2\right)-\dd t^2\label{2.27}\\[1.5ex]
&\varepsilon=-1: & \dd
s^2&=a^2(t)\left(\dd\chi^2+\sinh^2\!\chi\,\dd\varphi^2+\cosh^2\!\chi\,\dd
z^2\right)-\dd t^2\nonumber
\end{align}
and the mass density is (see \eqref{2.19})
\begin{equation}\label{2.28}
\mu=\frac{3a_\textrm{m}}{\kappa a^3(t)}\ .
\end{equation}
%%%%%%%%%%%%%%%%%%%%%%%%%%%%%%%%%%%%%%%%%%%%%%%%%%%%%%%%%%%%%%%%%%%%
\section{Interpretation of the solutions}

With the line elements \eqref{2.27} and the dynamical equation
\eqref{2.24} we finally arrived at the geometry
of the Friedman models, here described in double polar coordinates $\chi$,
$\varphi$, $z$, cf.~\cite{Koordinaten}. This is quite amazing since we
started with cylindrical symmetry and tried to obtain collapse
models different from the spherically symmetric Oppenheimer-Snyder
model.   

\begin{figure}
\begin{center}\resizebox{6.5cm}{!}{\includegraphics{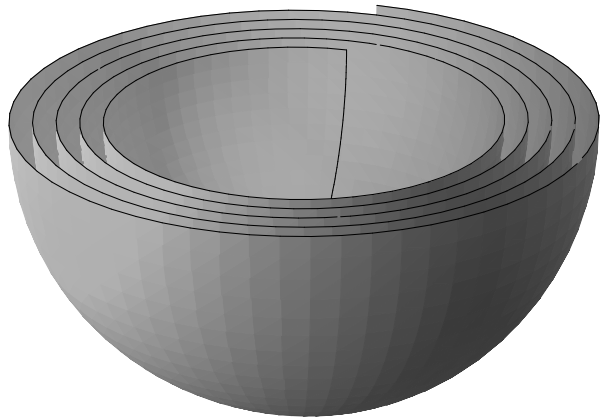}}\end{center}
%\begin{center}\resizebox{8cm}{6cm}{\includegraphics{Schalen7.eps}}\end{center}
\caption{Illustration of the solution's cylindrical interpretation}\label{fig1}
\end{figure}

In double polar coordinates the $z$-coordinate is limited by 
$z\in[0,2\pi]$. However, for a cylinder, $z$ should take on all real values.
Here, we provide a suitable topological interpretation to circumvent
this contradiction:
For $\varphi=\textrm{constant},
t=\textrm{constant}$,the  first line element in \eqref{2.27}, for example, reduces to
\begin{equation}\label{2.36}
\dd s^2=a^2(\dd\chi^2+\cos^2\!\chi\,\dd z^2)\ .
\end{equation}
Of course, this is the line element for a sphere of radius
$a$ with the angular variables $\chi$ and $z$. Now, instead of only one
sphere, we consider an infinite number of connected shells (as
sketched in fig. \ref{fig1}). 
If $z$ increases by $2\pi$ one moves from one skin
onto another.

This topology leads to a cylindrical interpretation of the solution.

%%%%%%%%%%%%%%%%%%%%%%%%%%%%%%%%%%%%%%%%%%%%%%%%%%%%%%%%%%%%%%%%%%%%
\section{discussion}
The intention of this paper was to find a cylindrically symmetric solution
for collapsing dust. It turned out that the assumption that the axial
and azimuthal metric functions fulfill eq. \eqref{2.3a}  leads to only one class of
solutions: collapsing Friedman Universes. The apparent contradiction
between the spherical symmetry of the Friedman models and the required
cylindrical symmetry was eliminated using a new topological
interpretation. Thus we have found an analytic solution describing a collapsing
infinite cylinder of dust. To have a complete solution of the Einstein
equations, one has to compute the exterior vacuum space-time that matches
to the given interior solution (if such a solution exists, which is not
clear a priori). For the vacuum region one should use the
line element
\begin{equation}\label{2.37}
\dd s^2=\e^{-2U}\left[\e^{2k}(\dd\rho^2-\dd
t^2)+\rho^2\dd\varphi^2\right]+\e^{2U}\dd z^2\ .
\end{equation}
Then one has to solve the field equations
\begin{equation}\label{2.38}
U''+\frac{1}{\rho}U'-\ddot U=0\ ,\quad k'=\rho(U'^2+\dot U^2)\ ,\quad
\dot k=2\rho U'\dot U\ .
\end{equation}

For matching a vacuum region the picture in FIG. \ref{fig1} changes:
Only in the interior region, $\chi\le\chi_0$, is the metric the Friedman
solution in double polar coordinates. So one has to cut off a
segment around the south pole of the hemisphere of FIG. \ref{fig1} and connect
the inner solution from 
there to the vacuum region.
    
Unfortunately, first numerical attempts seem to run into singularities.

%%%%%%%%%%%%%%%%%%%%%%%%%%%%%%%%%%%%%%%%%%%%%%%%%%%%%%%%%%%%%%%%%%%%

\begin{acknowledgments}
This work was funded in part by SFB/Transregio 7.
\end{acknowledgments}

%%%%%%%%%%%%%%%%%%%%%%%%%%%%%%%%%%%%%%%%%%%%%%%%%%%%%%%%%%%%%%%%%%%%

\end{document}